%% file: tmi.tex
\documentclass[journal,twoside,web]{IEEEtran}
\usepackage{cite}
\usepackage{amsmath,amssymb,amsfonts}
\usepackage{algorithmic}
\usepackage{graphicx}
\usepackage{textcomp}

\usepackage[svgnames,table]{xcolor}
\usepackage{float}
\usepackage{booktabs}
\usepackage{mathtools}
\usepackage{subcaption}
\usepackage{multirow}

\usepackage{soul}
\usepackage{caption, copyrightbox}

\def\BibTeX{{\rm B\kern-.05em{\sc i\kern-.025em b}\kern-.08em
    T\kern-.1667em\lower.7ex\hbox{E}\kern-.125emX}}
\begin{document}
\title{Geometry-aware neural solver for fast Bayesian calibration of brain tumor models}
\author{Ivan Ezhov, Tudor Mot, Suprosanna Shit, Jana Lipkova, Johannes C. Paetzold, Florian Kofler, Fernando Navarro, Chantal Pellegrini, Marcel Kollovieh, Marie Metz, Benedikt Wiestler and Bjoern Menze, \IEEEmembership{Member, IEEE}

\thanks{The first two authors contributed equally.}
\thanks{I. Ezhov and S. Shit are supported by the Translational Brain Imaging Training Network under the EU Marie Sklodowska-Curie programme (Grant ID: 765148). B. Menze, B. Wiestler and F. Kofler are supported through the SFB 824, subproject B12, by DFG through TUM International Graduate School of Science and Engineering, GSC 81, and by the Institute for Advanced Study, funded by the German Excellence Initiative. Fernando Navarro is supported by DFG-GRK 2274.}
\thanks{I. Ezhov, T. Mot, S. Shit, J. Paetzold, F. Navarro, F. Kofler, C. Pellegrini and M. Kollovieh are with the Department of Informatics, and with TranslaTUM - Central Institute for Translational Cancer Research, TUM, Munich, Germany (e-mail: ivan.ezhov@tum.de).}
\thanks{B. Wiestler and M. Metz are with the Neuroradiology Department of Klinikum Rechts der Isar, TUM, Munich, Germany}
\thanks{J. Lipkova is with Harvard Medical School, Brigham and Women's Hospital, Boston, United States}
\thanks{B. Menze is with the Department of Informatics, TranslaTUM, TUM, Munich, Germany, and Department of Quantitative Biomedicine of UZH, Zurich, Switzerland}}

\maketitle

\begin{abstract}
Modeling of brain tumor dynamics has the potential to advance therapeutic planning. Current modeling approaches resort to numerical solvers that simulate the tumor progression according to a given differential equation. Using highly-efficient numerical solvers, a single forward simulation takes up to a few minutes of compute. At the same time, clinical applications of tumor modeling often imply solving an inverse problem, requiring up to tens of thousands of forward model evaluations when used for a Bayesian model personalization via sampling. This results in a total inference time prohibitively expensive for clinical translation. While recent data-driven approaches become capable of emulating physics simulation, they tend to fail in generalizing over the variability of the boundary conditions imposed by the patient-specific anatomy. In this paper, we propose a learnable surrogate for simulating tumor growth which maps the biophysical model parameters directly to simulation outputs, i.e. the local tumor cell densities, whilst respecting patient geometry. We test the neural solver in a Bayesian model personalization task for a cohort of glioma patients. Bayesian inference using the proposed surrogate yields estimates analogous to those obtained by solving the forward model with a regular numerical solver. The near real-time computation cost renders the proposed method suitable for clinical settings. The code is available at https://github.com/IvanEz/tumor-surrogate.
\end{abstract}


\section{Introduction}
\label{sec:introduction}
\IEEEPARstart{S}{imulation} of brain tumor progression can provide complementary information to medical imaging for radiotherapy planning. As shown in \cite{ref_article3,ref_article7,ref_article11,ender_person_appl,Le_2017,subramanian2020multiatlas,scheufele2020automatic,scheufele2019image,zhang2017convolutional,petersen2019deep,elazab2018macroscopic,mang2020integrated,clatz2005realistic}, tumor modeling can be employed to define a personalized radio-treatment area using biophysical models to estimate the most likely directions of tumor cell infiltration instead of solely targeting tumor area visible on a scan. These methodologies mainly imply solving an inverse problem: finding the parameters of the biophysical tumor growth model resulting in a simulation output that best matches an empirical observation outlining the pathology.

Existing approaches for inverse tumor modeling resort to deterministic \cite{ref_article7,ender_person_appl,Jackson_2015} as well as probabilistic Bayesian \cite{ref_article10,ref_article8,Ezhov_2019} formalisms. All these approaches rely on excessive amount of forward simulations required for inferring patient-specific parameters. The number of simulations ranges from several thousand for approximate methods \cite{Le_2017,Ezhov_2019,ref_article10} to tens of thousands in case of fully Bayesian analysis \cite{ref_article11}. The forward brain tumor models are often based on the reaction-diffusion equation and are implemented using highly-efficient numerical solvers. In \cite{ref_article8}, authors employ the Lattice Boltzmann method which allows parallelized computing and takes ca. 80 seconds on a 60 core machine for simulating the pathology growth. In \cite{ref_article11}, the forward model is implemented by means of a multi-resolution adapted grid solver with a simulation time of 1-3 minutes using 2 cores. Despite the computational advances of the solvers, the minutes of a single forward model evaluation multiplied by thousands of forward integrations necessary for the inverse problem can result in weeks of total computing time. This constrains the testing of more elaborate tumor models (e.g., considering cell mixtures or multiple competing patho-physiological processes \cite{cristini2010multiscale}), and translation of the personalized radiotherapy planning into clinical practice \cite{ref_article3,ref_article7,ref_article11,ender_person_appl}. 

\begin{figure*}[!ht]
\includegraphics[width=1.0\textwidth]{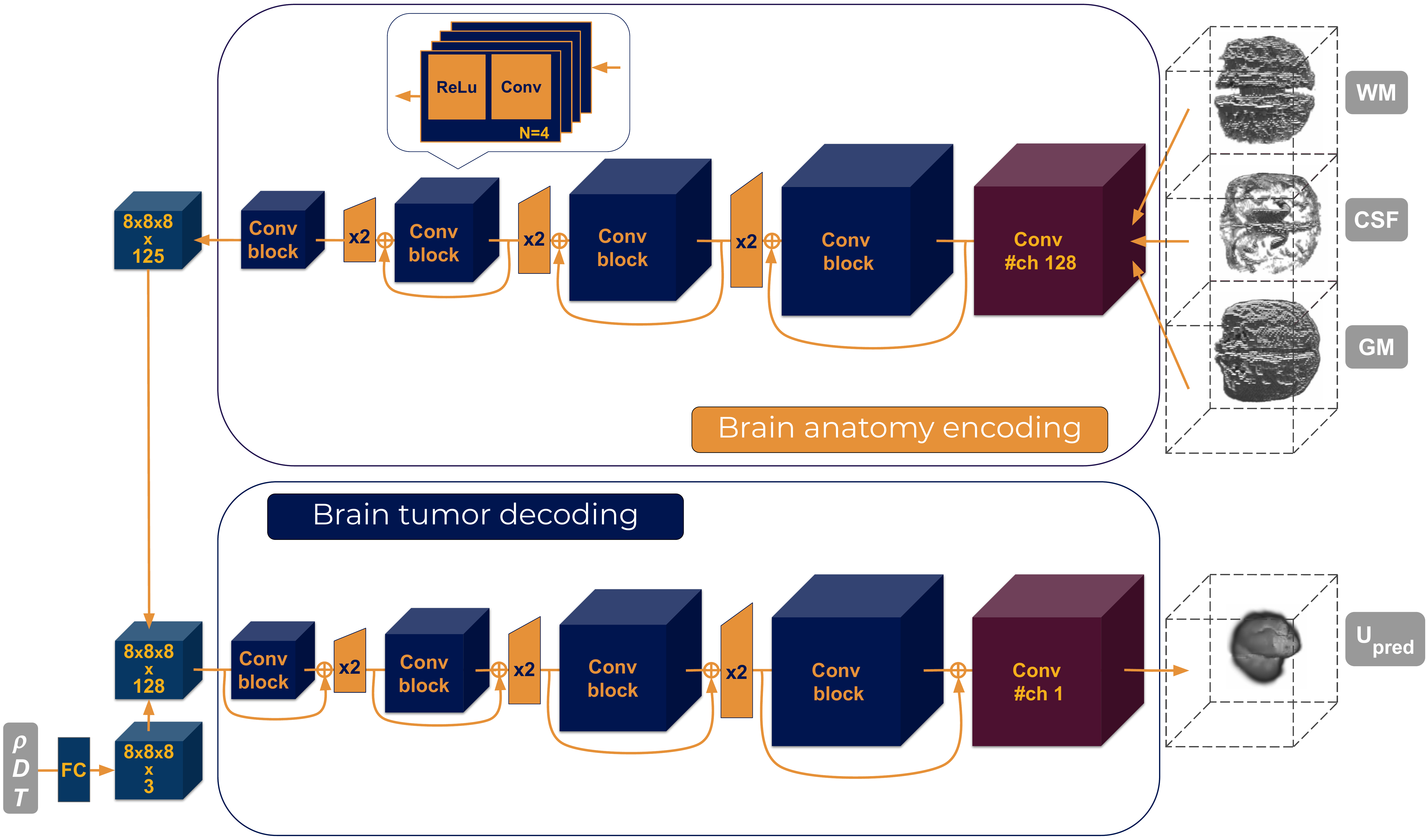}
\caption{\small{Geometry-aware neural solver. The network is composed of two main parts: a) brain anatomy encoder that maps the anatomy volumes (WM, GM, CSF) to a latent representation, b) brain tumor decoder that takes as input an embedding (via a fully-connected (FC) layer) of the parameters $\{D,\rho,T\}$, concatenated with the latent representation from (a), and maps the resulting tensor to the 3D tumor simulation volume. Downsampling in the encoder is implemented as a convolution with stride 2 (other convolutions in the network have kernel size 3 and stride 1). Upsampling in the decoder is via the nearest-neighbor interpolation followed by convolutional operations. The convolutional block is composed of N repetitions of convolutional operation (with number of channels $\#$ch 128) and ReLu non-linearity.}} \label{fig1}
\end{figure*}

As recent years showed, speeding up heavy conventional computation becomes feasible using end-to-end learning methods. The data-driven methodology has also penetrated the field of numerical computing \cite{dans1,dans2,dans4,dans5,dans6,dans7,dans8,dans9,dans10,geremia2012brain,kochkov}. Learnable surrogates were proposed for various scientific computing tasks in the natural sciences by exploiting fully-connected \cite{raissi2019physics,sitzmann2020implicit,stevens2020finitenet}, convolutional \cite{thuerey2020deep, kasim2020up,Kim_2019}, and hybrid \cite{Hsieh2019LearningNP,sht2019implicit,sanchez2020learning} neural architectures. Among them are two methods that proved capable of learning even a direct mapping from the space of parameters driving a simulator to the space of the simulator solutions in a static geometry \cite{kasim2020up, Kim_2019}. Unfortunately, these promising methods are incapable of dealing with inference in arbitrary complex geometries, such as those dictated by patient-specific anatomy. This limits their transfer to model personalization that is crucially dependant on an adaption to the patient specific simulation domain. 

The contribution of the paper is the following: we introduce a learnable method emulating a numerical tumor growth forward solver. Specifically, we introduce a learnable anatomy encoder that enables a patient-specific simulation of the tumor growth process. To the best of our knowledge, this is the first network-based approach in the computational pathology field that maps parameters of the biophysical model directly to the simulation outputs while generalizing over the simulation geometry. To illustrate the power of this approach, we use reaction-diffusion tumor growth model, since this is most used model and it also serves as a base for many more complex models. We achieve a 50$\times$ speed-up compared to an advanced numerical solver by employing the tumor model surrogate with an anatomy encoder that enforces patient-specific boundary conditions. This enables a fast Bayesian model personalization that is consistent with the baseline numerical solver.

\input{method}

\input{results}


\section{Conclusion}
We present the first learnable surrogate with anatomy encoder for tumor growth modeling that is capable of mapping the model parameters to the corresponding simulations while respecting patient-specific anatomy. Our method achieves real-time simulation 50$\times$ faster than numerical solver. Even though we tested  the  surrogate  on  the  simplistic  Fisher-Kolmogorov model, the technique can be readily adopted to more complicated tumor growth models and similar 4D inverse modeling problems.



\bibliographystyle{IEEEtran}
\bibliography{mybib}


\end{document}

%% file: method.tex
\section{Method}
\begin{figure}[]
\includegraphics[width=1.0\columnwidth]{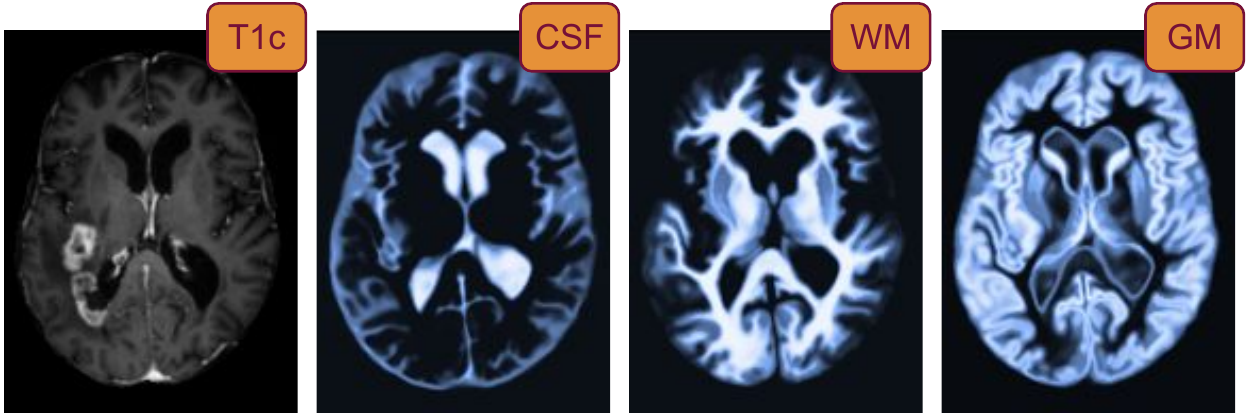}
\caption{\small{An example of an MRI T1c scan from the dataset and corresponding probabilistic segmentations offering information on the expected tissues underneath and nearby the tumor (obtained by registering the brain atlas to patient space).}} \label{fig6}
\end{figure}

\subsubsection{Forward tumor model}
The simulations that we aim to emulate are generated by a 3D numerical solver relying on the Fisher-Kolmogorov type of partial-differential equations (PDE). The equation describes the evolution of the pathology by considering diffusion and proliferation of the tumor cells under the Neumann boundary condition (B.C.):  
\begin{equation}
\frac{\partial \mathbf{u}}{\partial t}= \nabla (\textbf{D}\nabla \mathbf{u})+\rho \mathbf{u}(1-\mathbf{u}), \\
\end{equation}
\begin{equation}
\nabla \mathbf{u} \cdot \textbf{n} = 0 \quad B.C.
\end{equation}
Here, $\mathbf{u}$ is the normalized 3D tumor cell density, \textbf{D} denotes the diffusion tensor, $\rho$ is the rate of cell proliferation, and \textbf{n} is the normal vector to the boundary. We assume the infiltration of the tumor cells to occur only in white matter (WM) and grey matter (GM), considering isotropic diffusion with ${\textbf D} = D \cdot \mathbb{I}$ (where $D \in \{D_w,D_g\}$ is a diffusion coefficient in white or grey matter, and $\mathbb{I}$ is an identity matrix). Both WM and GM constitute the simulation domain while cerebrospinal fluid (CSF) and skull determine the patient-specific boundary. The segmentations are extracted from medical scans of patients diagnosed with the tumor. Given the probabilistic nature of the WM and GM segmentation maps used here, the diffusion coefficient is defined as $ D_{i}=p_{w_{i}} D_{w}+p_{g_{i}} D_{g}$, where $p_{w_{i}}$ and $p_{g_{i}}$ denote the percentage of the WM and GM at the voxel $i$. The diffusion coefficient in white matter Dw is considered
to be greater than in grey matter. We considered two models with the ratio $D_{w}/D_{g}$ equal 10 and 100. The input to the solver is a set of parameters $\theta_{P} = \{D_w,\rho,x,y,z,T\}$, where $x,y,z$ define the position of the function $\mathbf{u}$ at time $T=0$, which is initialized as a point source. 

\subsubsection{Learnable forward model surrogate with anatomy encoder}    
Our goal is to learn a tumor model surrogate which maps parameters of the pathology model $\theta_{P}$ to corresponding simulations $\mathbf{u}(\theta_P)$ given a patient-specific anatomy. We base our method on \cite{Kim_2019} which is designed to emulate the mapping for fluid simulations in a static spatial domain. Different from \cite{Kim_2019}, we need to consider patient-specific boundary conditions. To this end, we introduce an anatomy encoder that imposes anatomical boundary conditions, Fig. \ref{fig1}.

The numerical solver's output $\mathbf{u}(\theta_P)$ has a size of $128\times128\times128$ voxels. However, to provide a greater anatomical variability to the dataset on which we train the surrogate, we crop all simulated outputs and corresponding brain anatomies to $64\times64\times64$ volumes, centered at the initialization location $x,y,z$. The $64\times64\times64$ is greater than half the brain size and tumors bigger than this are incompatible with life.

The architecture of the tumor model surrogate consists of the following parts:

- \textit{Brain anatomy encoder} which encodes non-overlapping 3D volumes of the brain tissues WM, GM, and CSF through a series of convolutional blocks. The blocks are composed of alternating convolution operations (with fixed parameters of kernel size 3, stride 1, and the number of channels 128) and a non-linearity in the form of a linear rectifier. Each block is equipped with a skip connection linking the input and output of the block via an element-wise sum. Downsampling between the blocks is achieved by a convolutional operation with a stride of two,

- \textit{Brain tumor decoder} that takes a 1D vector of the parameters $\{D,\rho,T\}$ alongside with the encoded anatomy as input. Note that we do not condition the decoder on the initialization location $x,y,z$, since as mentioned above we crop all training volumes exactly at this location. Thus, the network is taught to reproduce the tumor in the center for any volume. Before being passed to the decoder, the 1D vector of the model parameters is mapped via a fully connected layer to a tensor of size $8\times8\times8\times3$ and is concatenated with the tensor of the encoded brain anatomy. The resulting tensor is gradually upsampled through a series of convolutional blocks analogous to the encoder and nearest-neighbor upsampling (we refer the reader to \cite{odena2016deconvolution} for a detailed discussion on why such type of upsampling is preferred over the deconvolution operation). At the end of the series, a 4D tensor $64\times64\times64\times128$ is convolved to the output prediction - a 3D tumor simulation volume (height 64, width 64, depth 64). The decoder design is adopted from \cite{berthelot2017began,Kim_2019}.

\begin{figure}[b]
\includegraphics[width=\columnwidth]{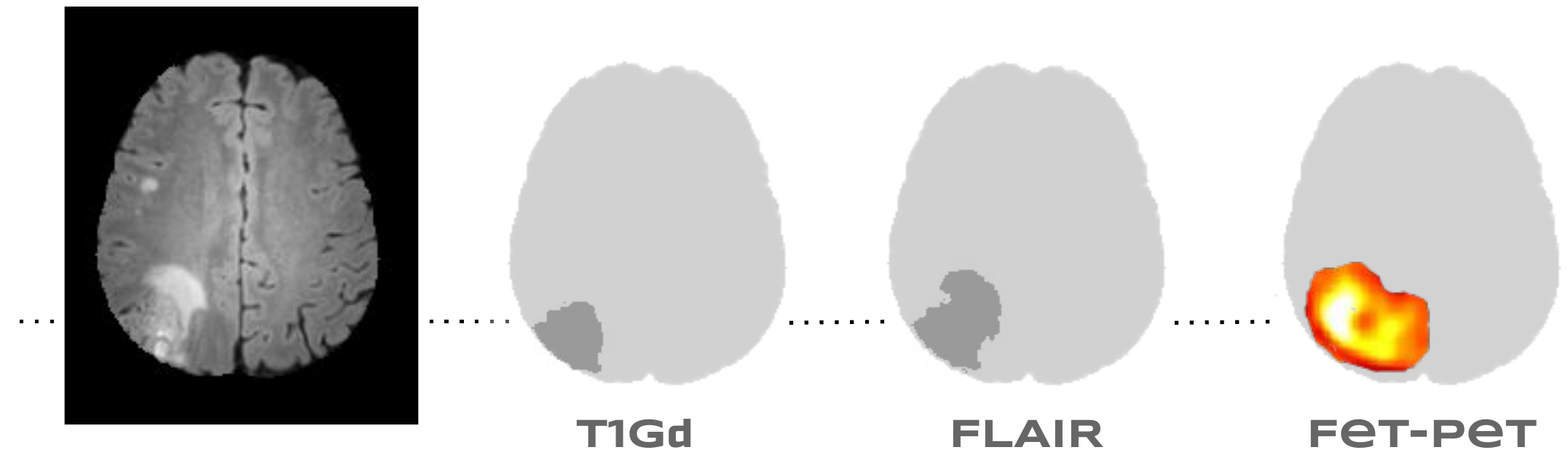}
\caption{An illustration of the imaging information used for Bayesian model calibration: binary segmentations obtained from T1Gd and FLAIR modalities, and FET-PET signal which is proportional to the tumor density.} \label{fig8}
\end{figure}

For constructing the \textit{loss function}, we can consider several semantically different sections of the output volume: CSF area (because no tumor cells are to be expected here), tumor area (because the algorithm should focus on this area ignoring the significantly larger normal brain), remaining brain area (composed of WM and GM voxels not present under tumor area), background (BG). We experimented with different combinations of this compartments in the loss, Tab. 1. The following loss definition outperforms other ways of defining spatial dependency in the cost function:
\begin{equation}
L_\mathbf{total}=\left\|u_{\mathbf{sim}}-{u}_{\mathbf{pred}}\right\|_{1}^{tumor}+\left\| u_{\mathbf{sim}}-u_{\mathbf{pred}}\right\|_{1}^{CSF}
\end{equation}
where the error between the predicted by the surrogate (${u}_{\mathbf{pred}}$) and simulated by the numerical solver ($u_{\mathbf{sim}}$) cell concentration is computed under $L_1$ norm separately in the \emph{tumor} area and in the \emph{CSF} area.

\begin{figure}[h]
\includegraphics[width=\columnwidth]{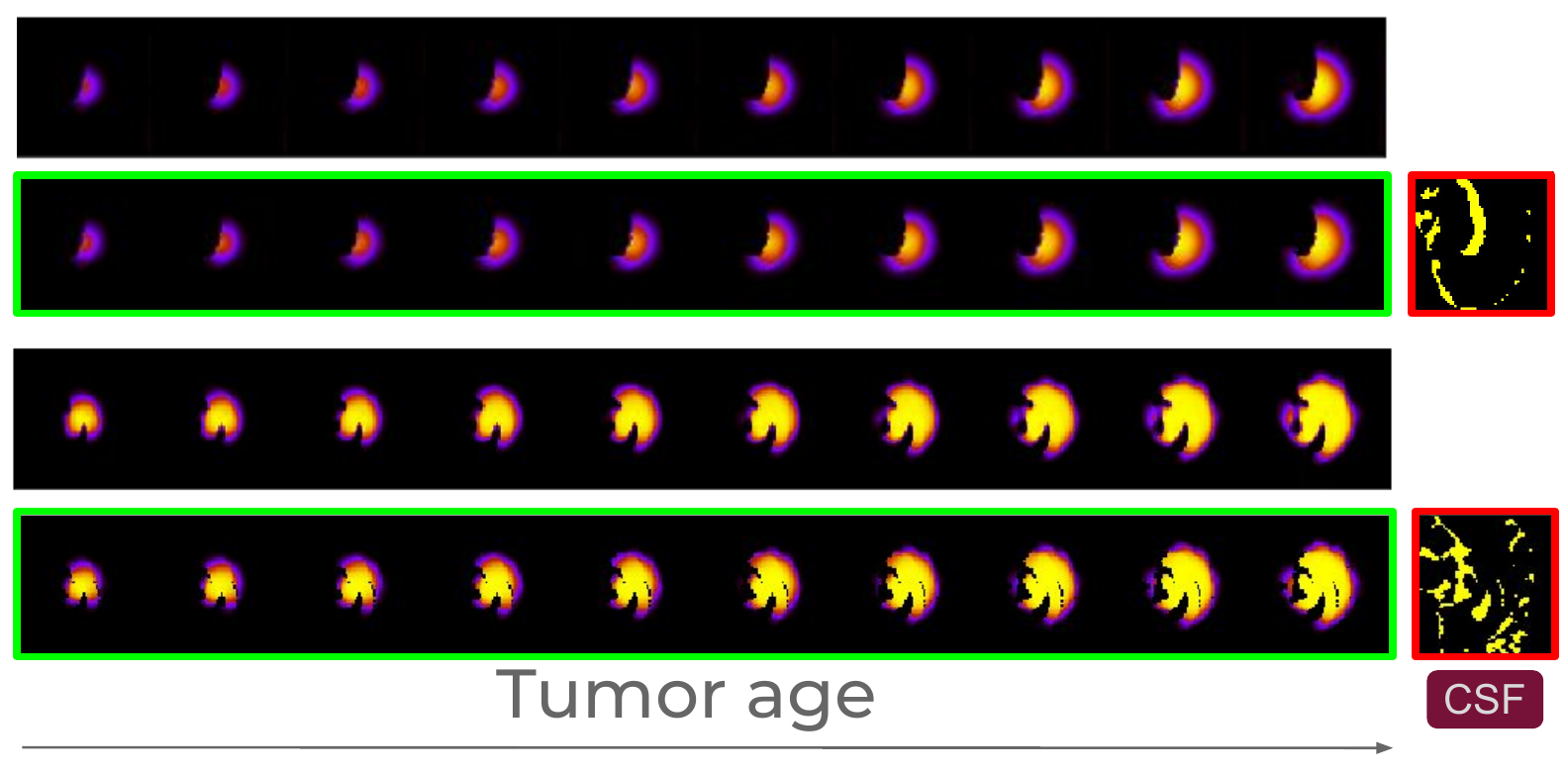}
\caption{Comparison between tumor volumes predicted by the surrogate and numerical solver. Each column represent a different value of tumor age $T$ from 100 to 1000 days with intervals of 100 days. The images were obtained by inferring individual 3D volumes for all time points and taking a central 2D slice from each volume. The rows framed in green correspond to the ground truth simulation. The CSF delineation constraining the tumor growth is framed in red.} \label{fig2}
\end{figure}

\begin{figure}[b]
\includegraphics[width=\columnwidth]{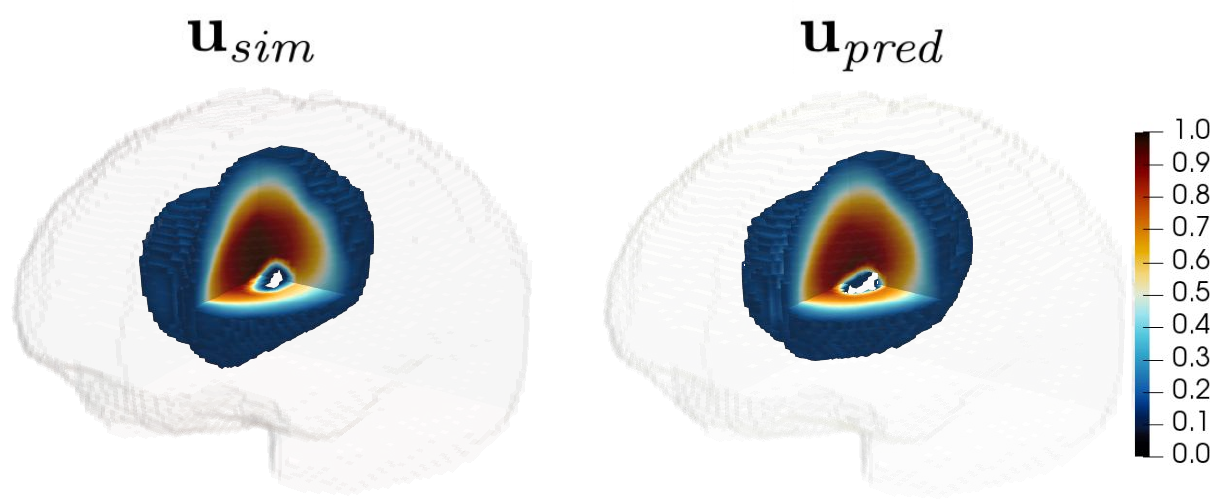}
\caption{A qualitative comparison between the simulated and predicted tumors embedded in the 3D brain anatomy.} \label{fig3}
\end{figure}

\begin{figure}[h]
\includegraphics[width=\columnwidth]{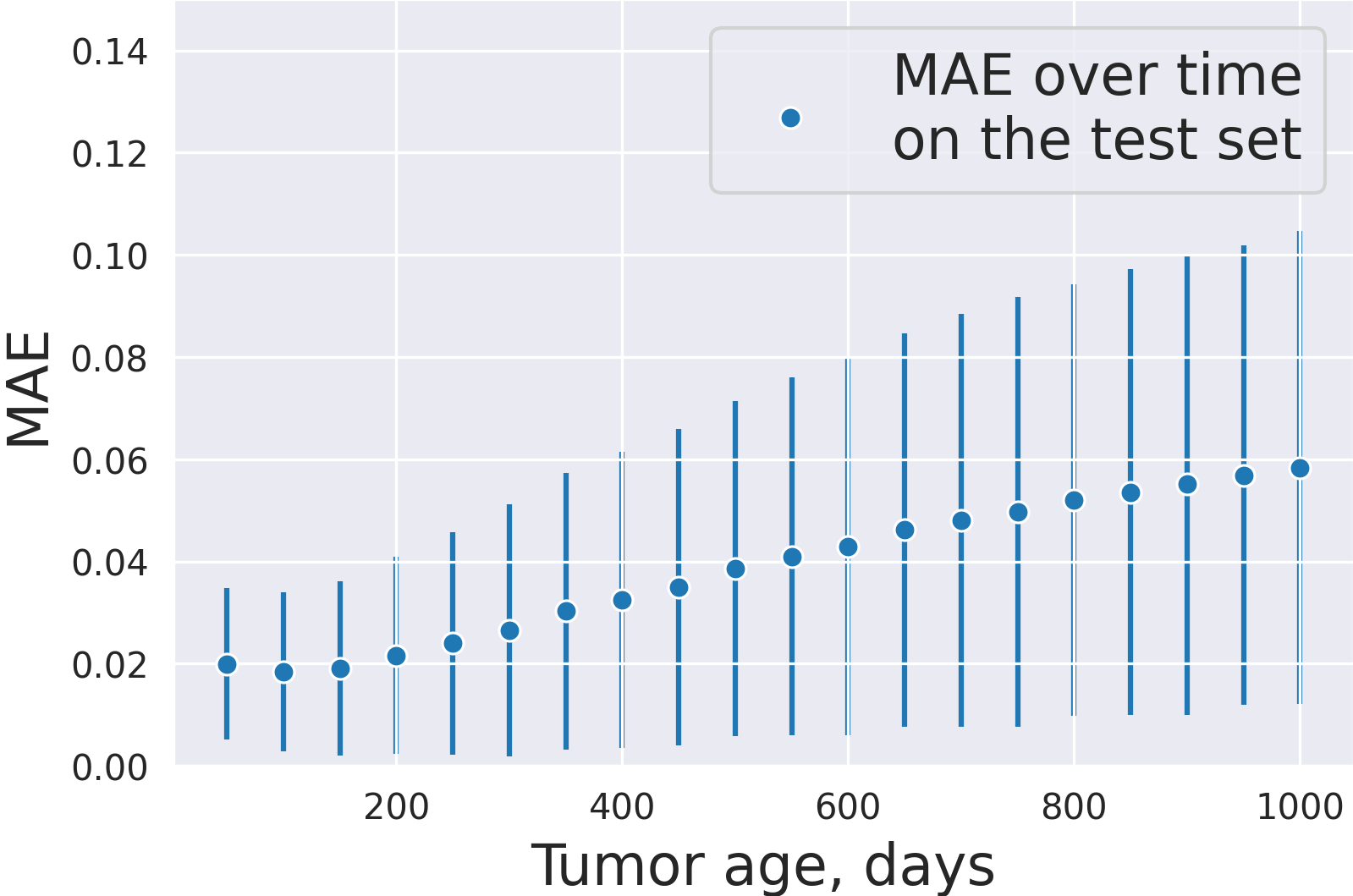}
\caption{The mean absolute error analysis over the tumor age (an input parameter) on the in-house test set. The error between the tumors (from numerical and neural solvers) is computed analogous to the first term in the proposed loss - in the area where the tumor concentration simulated using the numerical solver is greater than zero. Dots depict the means of the MAE distribution, error bars - the standard deviations. The analysis shows minor error  increase  with  the  increase  of  the  tumor  size.} \label{timeerrorfig}
\end{figure}

\subsubsection{Bayesian model personalization}

To demonstrate the applicability of the neural surrogate, we perform Bayesian tumor growth model personalization substituting the numerical solver with the learnable one. As calibration data we use two types of imaging modalities:
(a) T1 contrast-enhanced and FLAIR MRI modalities that allow estimating the morphological characteristic of the visible tumor; (b) FET-PET scans that provide information about the metabolic activity of the tumor.

Analogous to \cite{ref_article8,ref_article10}, we relate the output of the tumor growth solver $\mathbf{u}(\theta_P)$ to imaging information via a probabilistic model,

\begin{equation}
\begin{aligned}
   {p}(\mathcal{D} \mid \mathbf{u},\theta)={p}\left(\mathbf{y}^{T1c} \mid \mathbf{u},\theta_{I} \right) & \cdot {p}\left(\mathbf{y}^{F L A I R} \mid \mathbf{u},\theta_{I}\right)\cdot\\
   &\cdot {p}\left(\mathbf{y}^{P E T} \mid \mathbf{u},\theta_{I}\right)
\end{aligned}
\end{equation}

\noindent where image observations $\mathcal{D} = \{\mathbf{y}^{T1c}, \mathbf{y}^{{FLAIR}}, \mathbf{y}^{{PET}}\}$ are assumed to be independent, and $\theta = \{\theta_{P}, \theta_{I}\}$ constitute parameters of the pathology model $\theta_{P}$ and the probabilistic imaging model $\theta_{I}$. The latter is defined differently according to the type of modality:

\textit{- MRI modalities} provide information in the form of binary tumor segmentations ($\mathbf{y}^{T1c}, \mathbf{y}^{FLAIR}$). Thus we assign for each voxel a discrete label $y_{i}^{M} \in \{0,1\}$, $M \in \{T1c, FLAIR\}$. We model the probability of observing $\mathbf{y}^{T1c}, \mathbf{y}^{FLAIR}$ for a given concentration $\mathbf{u}(\theta_P)$ with Bernoulli distribution,

\begin{equation}
\begin{aligned}
p\left(\mathbf{y}^{T1c, FLAIR} \mid \mathbf{u}, \theta_{I}^{M}\right)&=\prod_{i} p\left(y_{i}^{M} \mid {u_i}, \theta_{I}^{M}\right)=\\
&=\prod_{i} \alpha_{i,M}^{y_{i}^{M}} \cdot\left(1-\alpha_{i,M}\right)^{1-y_{i}^{M}},   
\end{aligned}
\end{equation}

\noindent where $\alpha_{i,M}$ is the probability of observing tumor-induced changes defined as a double logistic sigmoid,

\begin{equation}
    \alpha_{i,M}\left(u_{i}, u_{c}\right)=0.5+0.5 \cdot \operatorname{sign}\left(u_{i}-u_{c}^{M}\right)\left(1-e^{-\frac{\left(u_{i}-u_{c}^{M}\right)^{2}}{\sigma_{\alpha}^{2}}}\right)
\label{eq_mri}
\end{equation}

\noindent With this formulation, we postulate that the tumor is not visible on a scan below the threshold level $u_c^M$. The parameter $\sigma_{\alpha}$ is introduced to take uncertainty in the threshold concentration $u_c^M$ into account. 

\begin{table*}
\centering
\begin{tabular}{*{10}c}
&  \vline   &  & Validation set & & \vline & & Test set &  \\\midrule
&  \vline   &  \textbf{$u_c=0.2$} & \textbf{$u_c=0.4$} & \textbf{$u_c=0.8$} & \vline & \textbf{$u_c=0.2$} & \textbf{$u_c=0.4$} & \textbf{$u_c=0.8$} \\\midrule

& \multicolumn{4}{c}{\textit{Ablation analysis on the network structure}} \\\midrule

\begin{tabular}[c]{@{}c@{}}(a) With skip-connections (U-net) \end{tabular} &  \vline   & 0.572 & 0.575 & 0.576 &  \vline   & 0.563 & 0.565 & 0.569 \\

\begin{tabular}[c]{@{}c@{}}(b) Only CSF as input \end{tabular} &  \vline & 0.739 & 0.734 & 0.734 &    \vline  & 0.768 & 0.767 & 0.763 \\\midrule

& \multicolumn{4}{c}{\textit{Ablation analysis on the loss function}} \\\midrule

\begin{tabular}[c]{@{}c@{}}(c) [Tumor+CSF+Brain+BG] \end{tabular} &  \vline   & 0.615 & 0.607 & 0.585 &  \vline   & 0.588 & 0.571 & 0.544 \\

\begin{tabular}[c]{@{}c@{}}(d) Tumor+[CSF+Brain+BG] \end{tabular} &  \vline   & 0.755 & 0.761 & 0.762 &  \vline   & 0.787 & 0.782 & 0.778 \\

\begin{tabular}[c]{@{}c@{}}(e) Tumor+CSF+Brain \end{tabular} &  \vline & 0.771 & 0.764 & 0.762 &  \vline   & 0.791 & 0.786 & 0.780 \\
\rowcolor{LightBlue!40}
\begin{tabular}[c]{@{}c@{}}(f) Tumor+CSF\end{tabular} &  \vline & 0.796 & 0.783 & 0.788 &  \vline   & 0.802 & 0.795 & 0.794 \\
\hline
\end{tabular}
\caption{Ablation analysis on the in-house validation (2k samples) and test (10k samples) tests: a) with skip connections between the encoder and decoder (U-Net like); b) instead of inputting 3 volumes of different tissue types (WM, GM, CSF) only a single volume of CSF tissue serves as an input to the network; c) a single loss is computed for the whole volume; d) the second term in the loss is computed in the spatial complement of the first \emph{tumor} term in Eq. 3; e) in addition to the two terms in Eq. 3, a third term is added to penalize for false predictions in the remaining brain, f) the proposed architecture. The values represent the means of the DICE score histograms. The DICE is computed for the tumor volumes $\mathbf{u}_{\mathit{pred}}$ and $\mathbf{u}_{\mathit{sim}}$ thresholded at 0.2, 0.4, and 0.8 values of tumor cell concentration. The square brackets denote the areas included in a single loss term, e.g. for [CSF+Brain+BG] the term is computed in the volume combining all three compartments.}
\end{table*}

\textit{- FET-PET modality} ($\textbf{y}^{PET}$) provides continuous information in each voxel and can be assumed to be proportional to the tumor density \cite{stockhammer2008correlation,hutterer201318f,ref_article11}. In this case we model the likelihood imaging function by a Gaussian distribution with unknown constant of proportionality $b$:
\begin{equation}
    p\left(\mathbf{y}^{PET} \mid \mathbf{u}, \theta_{I}^{M}\right)=\prod_{i} p\left(y_{i} \mid u_i, \theta_{I}^{M}\right)=\prod_{i} \mathcal{N}\left(y_{i} - bu_{i}, \sigma^{2}\right)
\end{equation}

\noindent Here, $\sigma$ models uncertainty in the PET signal, which is considered to be normalized $y^{PET}_i \in [0, 1]$. 

In total there are eleven parameters (six pathology model parameters $\theta_{P}=\{D_w,\rho,x,y,z,T\}$ and five imaging model parameters $\theta_{I}=\{u^{T1c}_c,u^{FLAIR}_c, \sigma_{\alpha},b,\sigma\}$) which we infer from the triplet of medical scans $\mathcal{D} = \{\mathbf{y}^{T1c}, \mathbf{y}^{{FLAIR}}, \mathbf{y}^{{PET}}\}$ using a Markov Chain Monte Carlo (MCMC) sampling algorithm \cite{ching2007transitional}.

\begin{figure*}[]
\captionsetup[subfigure]{labelformat=empty}
\begin{subfigure}{0.33\textwidth}
\includegraphics[width=\linewidth]{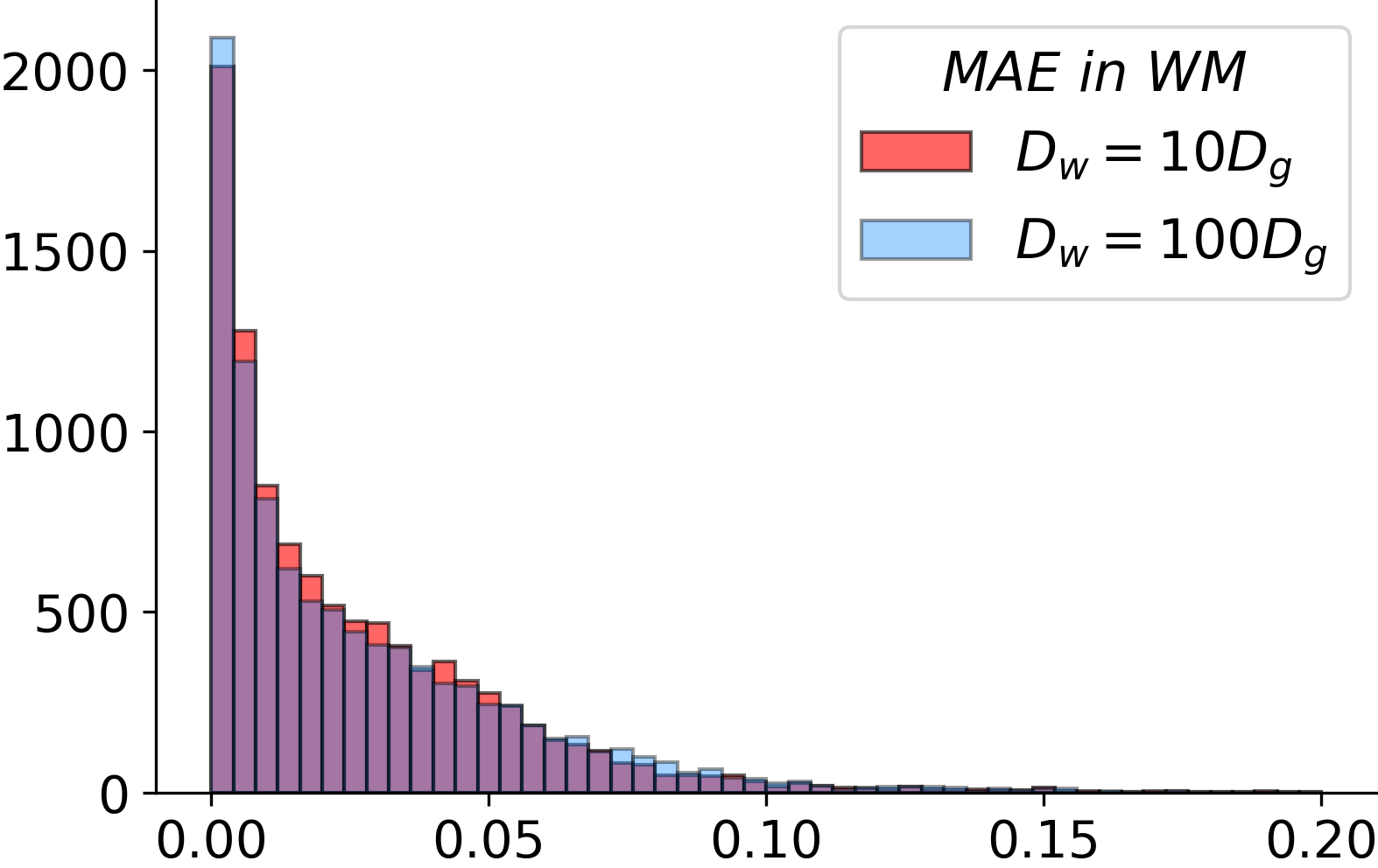}
\caption{MAE}
\end{subfigure}%
\begin{subfigure}{0.33\textwidth}
\includegraphics[width=\linewidth]{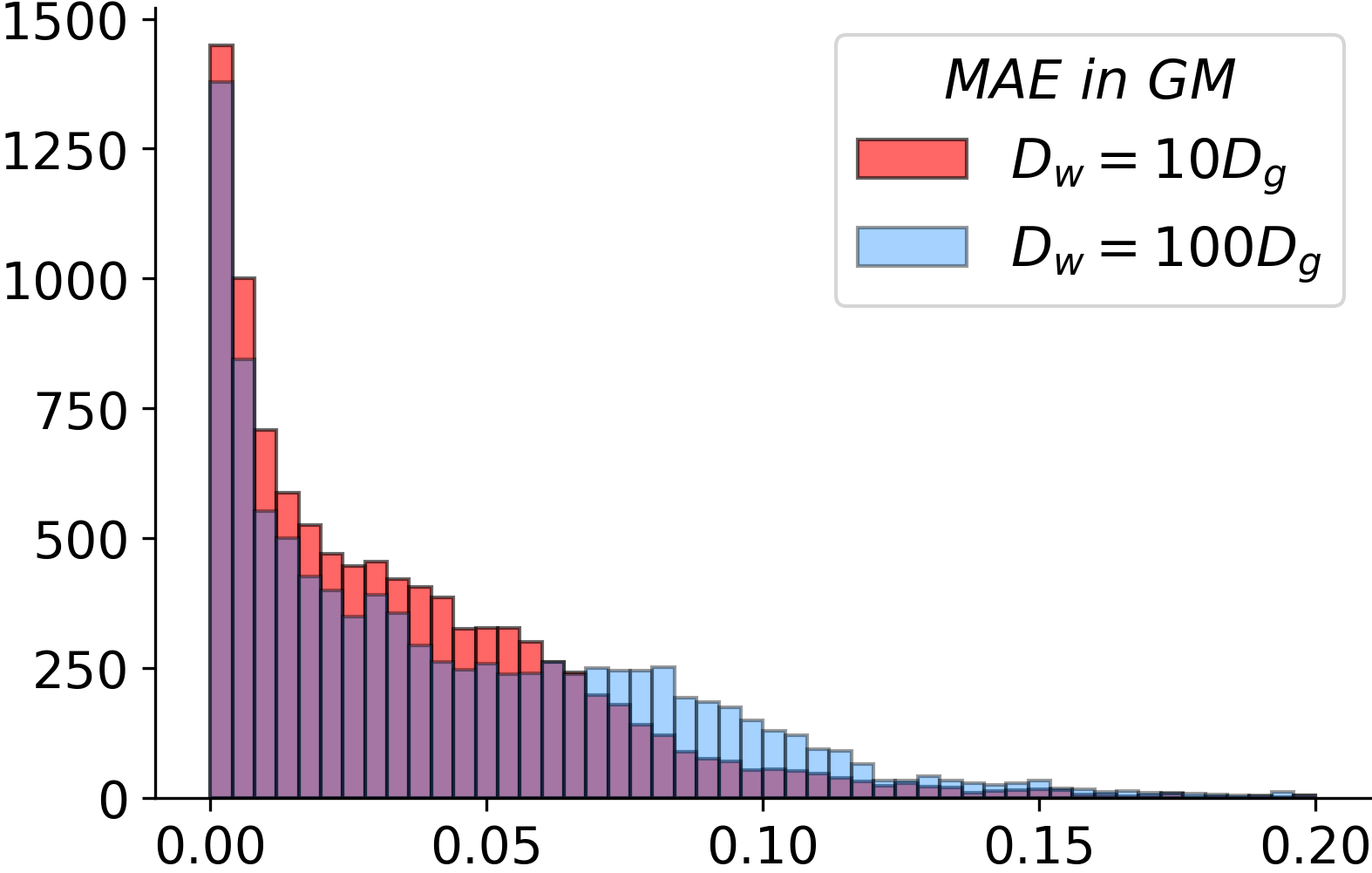}
\caption{MAE}
\end{subfigure}%
\begin{subfigure}{0.33\textwidth}
\includegraphics[width=\linewidth]{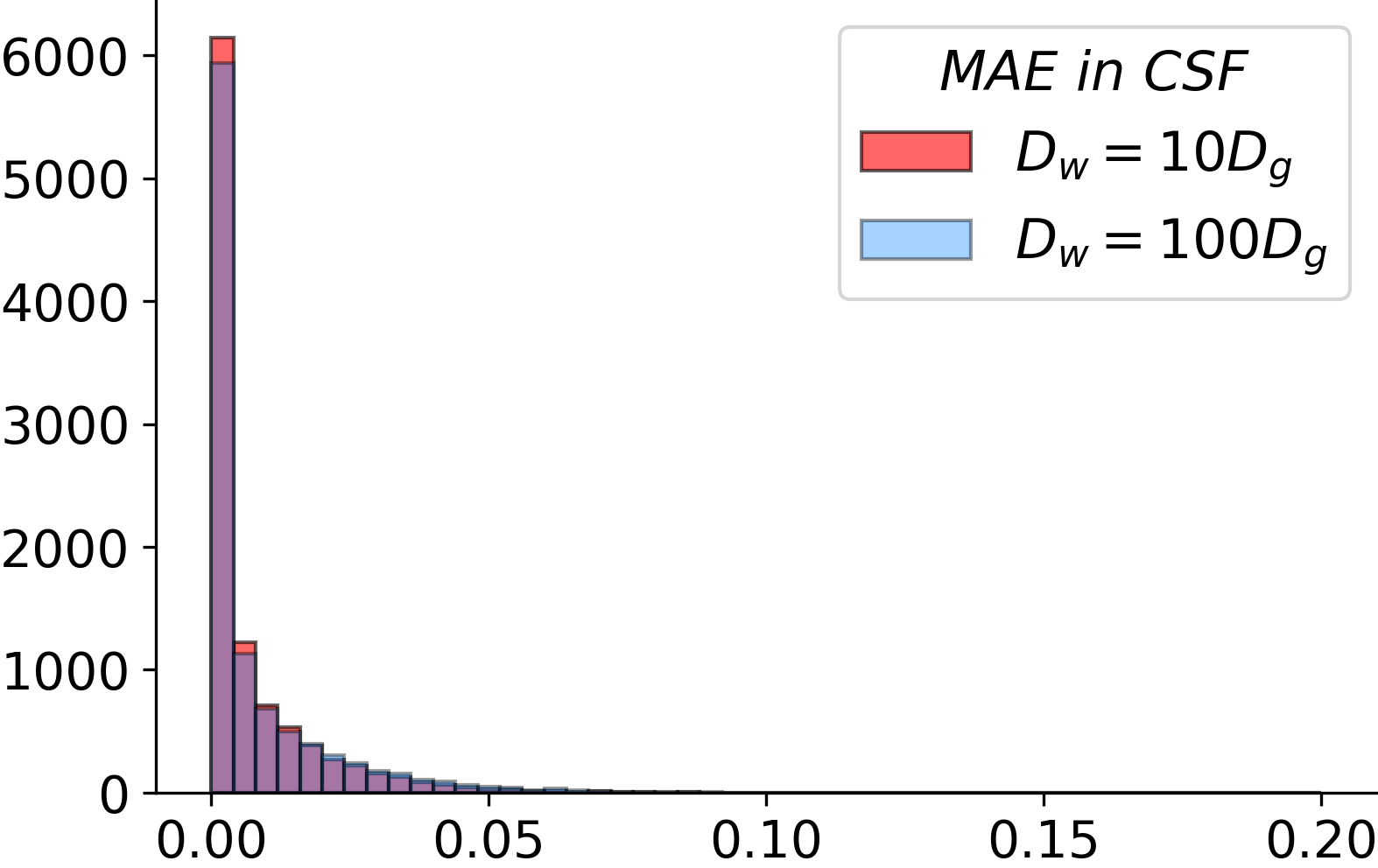}
\caption{MAE}
\end{subfigure}%
\caption{Histograms of the mean absolute error $\|\mathbf{u}_{\mathit{pred}}-\mathbf{u}_{\mathit{sim}}\|_{1}$ computed on the in-house test set (10k samples) within each class of the brain tissues (WM, GM, CSF). In red, samples from the tumor model with $D_w=10D_g$, in blue - from the tumor model with $D_w=100D_g$. The means of the distributions for the $D_w=10D_g$ model are: 0.026 (WM), 0.035 (GM), 0.007 (CSF). The means for the $D_w=100D_g$ model are : 0.028 (WM), 0.044 (GM), 0.008 (CSF). The fact that the network performance on two tumor models with notably different diffusion character stays close suggests that the network is capable of capturing not only pixel-wise CSF dependency, but also highly non-local diffusion dependency on the  WM  and GM.} 
\label{fig10}
\end{figure*}

\begin{figure*}[]
\captionsetup[subfigure]{labelformat=empty}
\begin{subfigure}{0.33\textwidth}
\includegraphics[width=\linewidth]{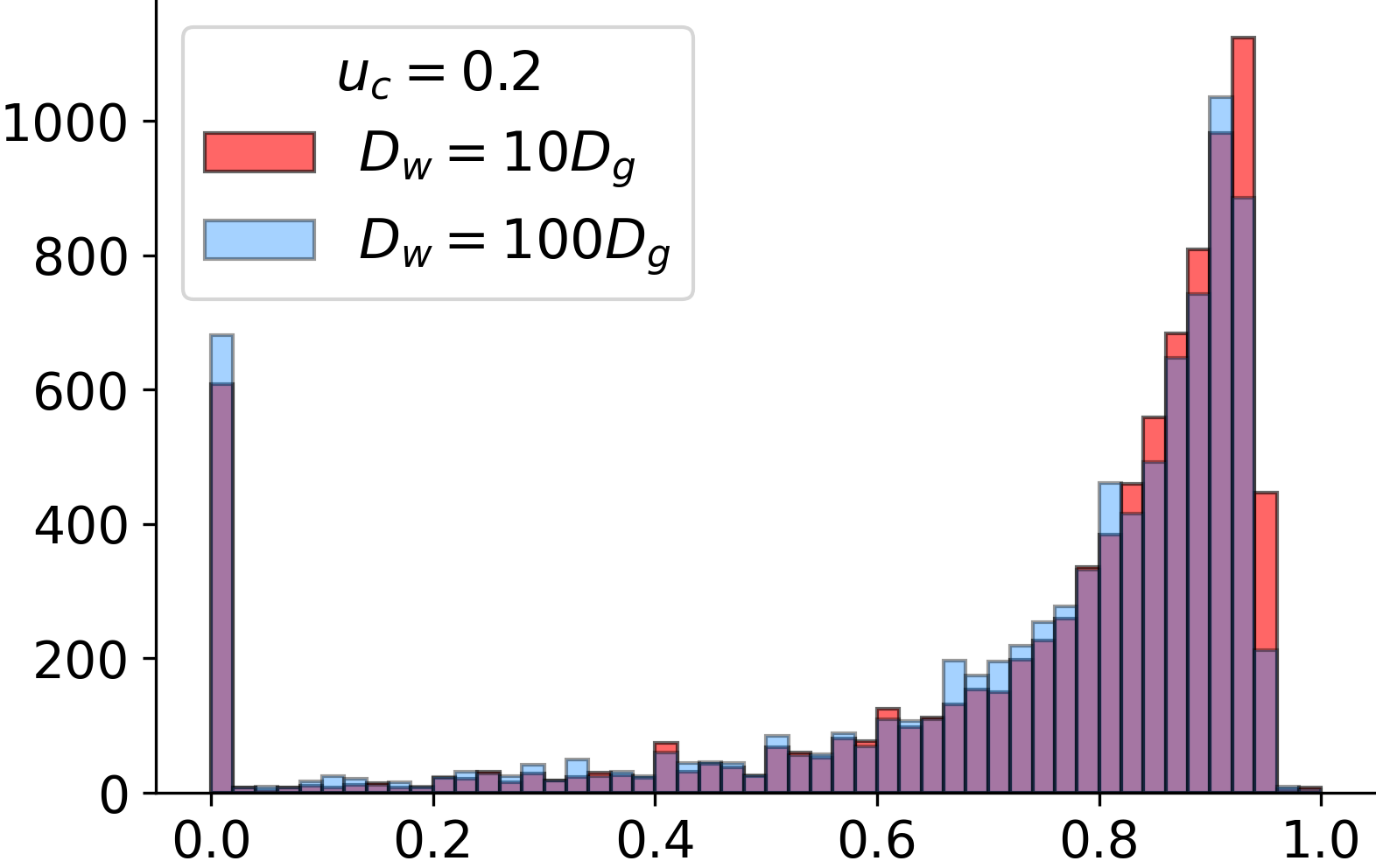}
\caption{DICE}
\end{subfigure}%
\begin{subfigure}{0.33\textwidth}
\includegraphics[width=\linewidth]{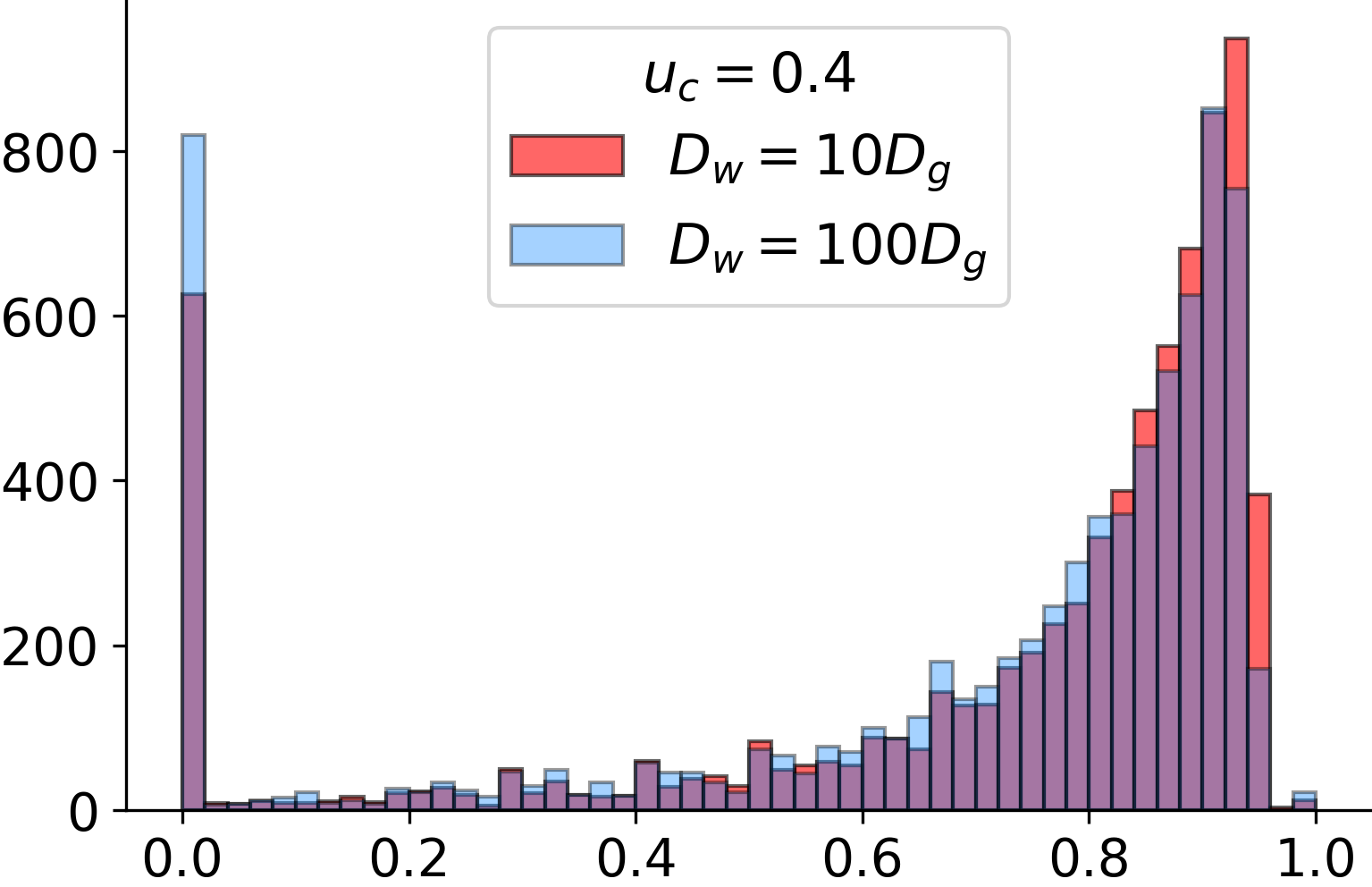}
\caption{DICE}
\end{subfigure}%
\begin{subfigure}{0.33\textwidth}
\includegraphics[width=\linewidth]{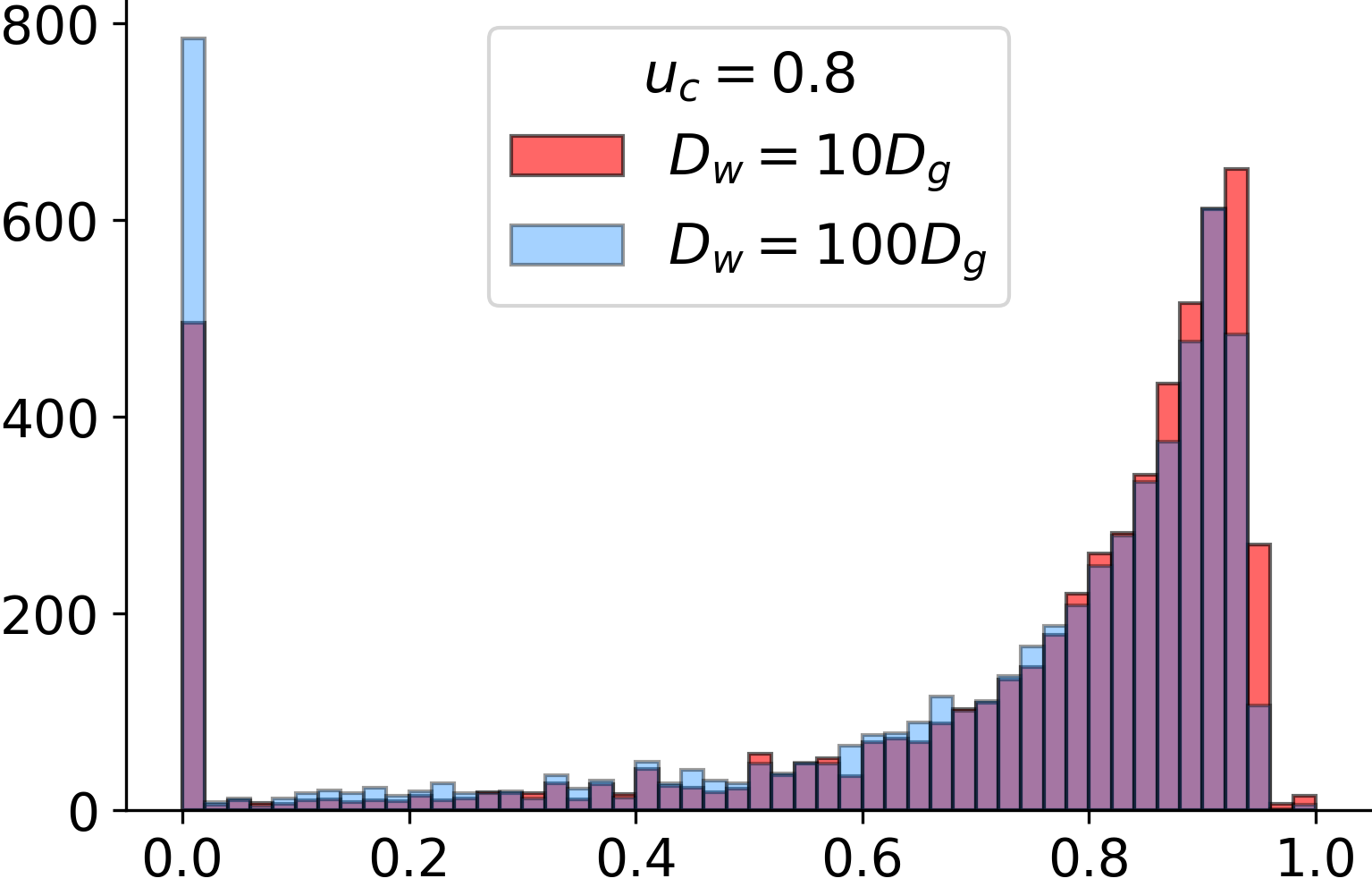}
\caption{DICE}
\end{subfigure}%
\caption{Histograms of the DICE score computed on the test set (10k samples) for the tumor volumes $\mathbf{u}_{\mathit{pred}}$ and $\mathbf{u}_{\mathit{sim}}$ thresholded at 0.2, 0.4, and 0.8 values of tumor cell concentration. In red, samples from the tumor model with $D_w=10D_g$, in blue - from the tumor model with $D_w=100D_g$. The means of the distributions for the $D_w=10D_g$ model are: 0.802 ($u_c=0.2$), 0.795 ($u_c=0.4$), 0.794 ($u_c=0.8$). The means for the $D_w=100D_g$ model are: 0.782 ($u_c=0.2$), 0.775 ($u_c=0.4$), 0.764 ($u_c=0.8$).} 
\label{fig11}
\end{figure*}

\subsubsection{Implementation}
The numerical tumor solver used for obtaining the simulation dataset is a highly parallelized glioma solver returning a 3D normalized tumor concentration profile on a uniform spatial grid \cite{ref_article11}.

The surrogate network is trained using the Adam optimizer with decay rates $\beta_1=0.5$ and $\beta_2=0.999$ for 30 epochs, which was observed to be sufficient for convergence. The learning rate is cosine annealed from $10^{-4}$ to $2.5\times10^{-6}$ over the training and the batch size is 16. The input parameters' ranges are min-max normalized. The training time is about 6 days on an NVIDIA Quadro P8000 using the Tensorflow framework.

\begin{figure}
\includegraphics[width=1.0\columnwidth]{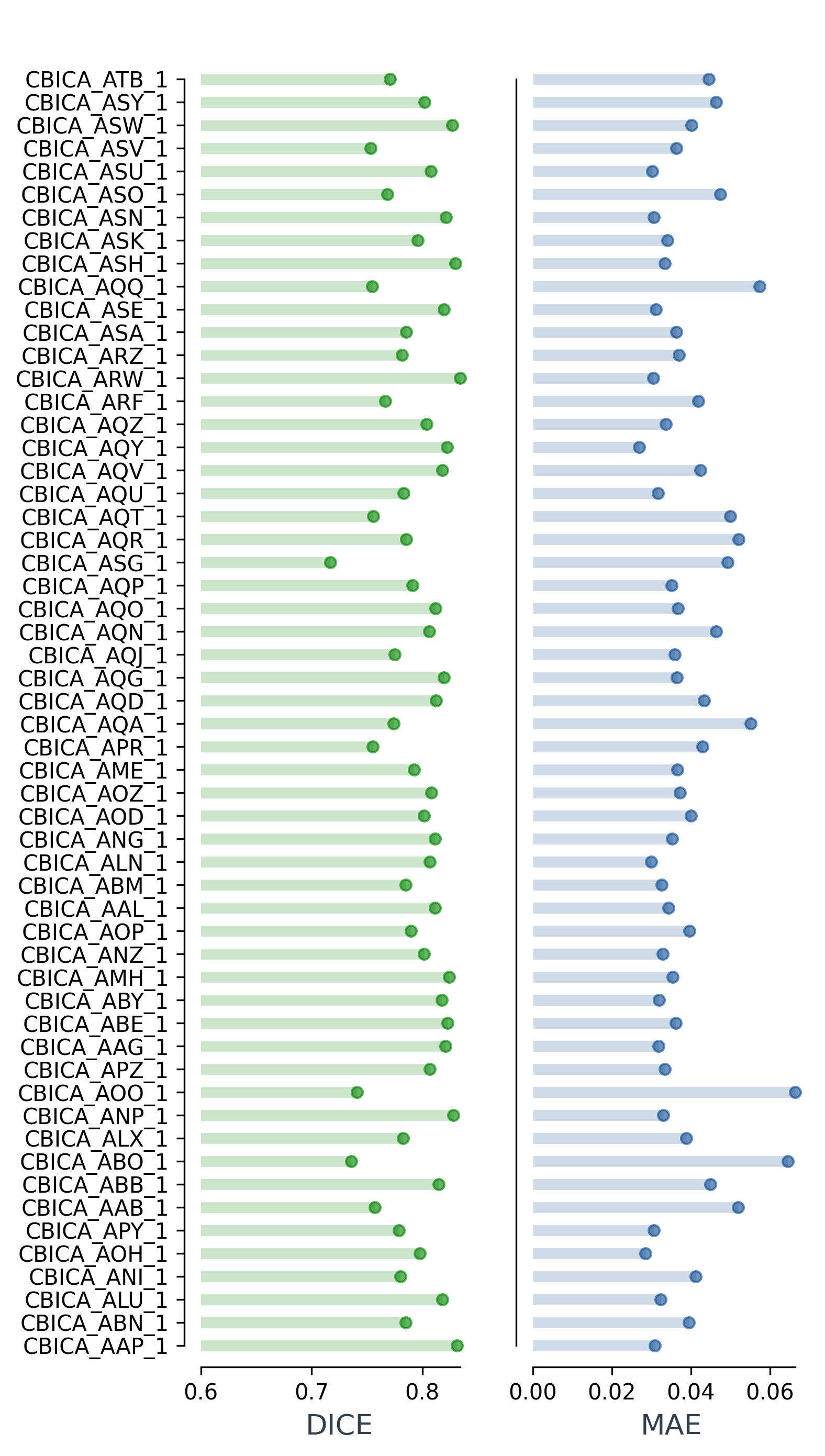}
\caption{\small{Distribution of the per-patient mean errors between the tumors predicted by the numerical and neural solvers over 56 patients from the BraTS dataset. The computed errors are DICE (thresholded at $u_c=0.2$) and MAE. For each patient the mean DICE and MAE are computed on 1k tumor simulations. Such evaluation on an independent test set demonstrates that the proposed neural surrogate possesses a high generalization ability that is stable across a high number of tumor and patient anatomies.}} \label{figbrats} 
\end{figure}

For Bayesian MCMC inference we use an implementation of Transitional MCMC from \cite{hadjidoukas2015pi4u} with 2048 samples per iteration. When the MCMC is used with the numerical solver, simulations of $128\times128\times128$ size are calibrated against the imaging information of the corresponding size. In the case of the surrogate, the network  predicts $64\times64\times64$ tumor volumes,  which  are  then  embedded  in  the  original $128\times128\times128$ domain for the calibration.

%% file: results.tex
\section{Experiments}
Conventional application of tumor modeling such as model personalization (via solving an inverse problem) implies estimating the model parameters by sampling over fixed, physiologically plausible ranges. Thus, we aim to employ the learnable surrogate which i) possesses an interpolation capacity for the parameters $\{D,\rho,T\}$, and ii) is capable to extrapolate for simulations in new geometries. To probe these properties, the validation and test sets were formed to have only the brain anatomies unseen by the network during training, while the parametric $\{D,\rho,T\}$ triplets were sampled from the same ranges as for the training.

\subsubsection{Data and parameterization}
We simulate the tumors in probabilistic brain tissue (WM, GM, CSF) segmentations that are obtained by registering a patient scan to the healthy brain atlas \cite{ref_article11,kofler2020brats} (available at github.com/JanaLipkova/s3). Different from a plain segmentation of the patient anatomy directly from an MRI scan, such procedure offers an estimate of the anatomy underneath the tumor. Fig.2 shows an example of such an approximation. 
We use two datasets: a) \textit{in-house set} composed of MRI and PET images from 20 patients (simulations in 10 patients' anatomies are used as training data, simulations in additional 5 patients for validation, and in the remaining 5 patients for test), b) all CBICA 56 patients of high-resolution MRI from \textit{BraTS} \cite{menzebrats} (56k simulations are used only for test). Both datasets have resolution of isotropic voxels of 1mm side-length, but for simulation purposes, the calibration is performed in data downsampled to 128x128x128 that corresponds to 2mm per side-length. 

For most of experiments, we considered a tumor model in which the diffusion in white matter $D_w$ is greater than in grey matter $D_g$ by 10 times (except Fig. \ref{fig10} and \ref{fig11}, where two diffusion models are compared - with the ratio of 10 and of 100 between the coefficients). The following ranges are used for random uniform sampling of the model parameters: $D_w \in [0.01,0.08]$ $mm^2/day$, $\rho \in [0.0001,0.03]$ $1/day$, and $T \in [50,1000]$ $day$ with a step size of 50 days. The tumor location coordinates $x,y$, and $z$ are sampled uniformly within the brain volumes. It should be noted that this uniform prior can be modified to non-uniform distributions, such as those reported in \cite{esmaeili2018direction,ref_article3}, to increase sampling for locations with higher tumor incidence. The initial locations $\{x, y, z\}$ were samled within the WM and GM only. For samples with initial location closer than 32 pixels to any of the volume borders, we do extra padding when cropping to 64x64x64 size. In total, we have a set of 20k parameters-simulation pairs for training.  

\subsubsection{Identifying the optimal algorithm on in-house dataset}

We perform several experiments on the in-house dataset to identify an optimal architectural setup:

\paragraph{Experiments on the loss function} 
As mentioned in the method section the predicted 3D image can be decomposed into several semantic parts (CSF, tumor, WM/GM, background). An ablation analysis demonstrates that the highest performance is achieved using the loss that focuses solely on the tumor and CSF areas, Tab. 1f. Notably, such loss is superior despite absence of penalization in rest of the volume. 

\begin{figure}[h]
\includegraphics[width=\columnwidth]{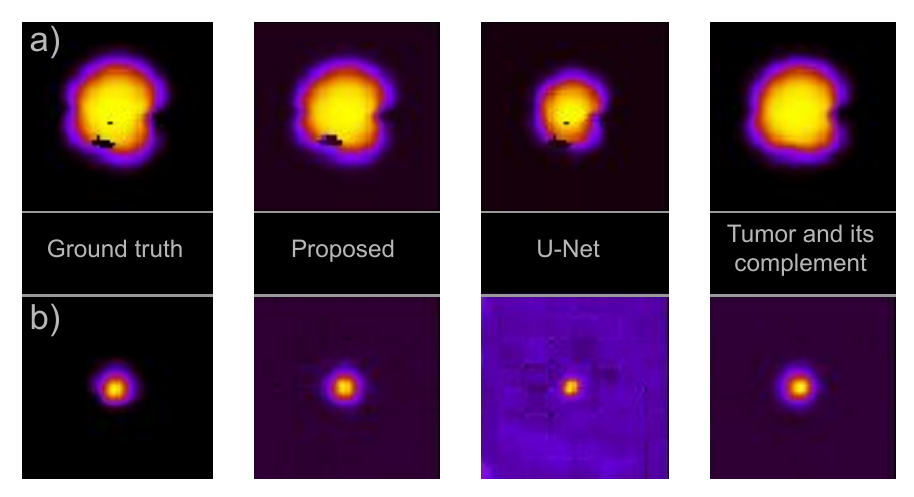}
\caption{Visual comparison of the inference obtained by the proposed method, U-Net, and the architecture with two terms in the loss - tumor and tumor complement (as in "d" in Tab. 1). The U-net exhibits mispredictions of two types: a) better CSF preservation is traded off for worse capturing of the parametric dependency, b) more notable false predictions in the out-of-tumor area. The proposed architecture can also produce false predictions in the non-CSF area (which is expected due to absence of related penalization in the loss). However, using penalization for the tumor complement area does not warrant absence of such noise ("Tumor and complement").} \label{unet}
\end{figure}

\paragraph{Experiments on the network structure} 
These days in the field of medical imaging computing, the natural baseline is a U-Net. We performed experiments comparing the U-net type architecture with the proposed encoder-decoder design. Fig. \ref{unet} depicts the problem with the U-Net, namely, worse capturing of the output volume's dependency on the model parameters, and more pronounced false predictions in the non-tumor area.

We also analyzed how much information the proposed network is capable to extract from the WM/GM signal. Results from Tab.1 show that despite the highly non-local character of dependency of the output tumor volume on the WM/GM, the performance is dropped by up to 6$\%$ if the WM/GM is not included in the input.  

\paragraph{Experiments on robustness} We perform experiments comapring two diffusion models with notably different diffusion character ($D_w/D_g=10$, $D_w/D_g=100$). Fig. \ref{fig10} demonstrates the distribution of the mean absolute error within each class of the brain tissues evaluated on the whole in-house test set (10K samples) for two types of diffusion models. Even though we observe samples with the error in the order of $10^{-1}$, the majority of the distribution lies within the order of $10^{-2}$. In Fig. \ref{fig11} we depict the histograms of the DICE score ($2|X\cap Y|/(|X|+|Y|)$) computed between the simulated and predicted tumor volumes which are thresholded at different levels of the tumor cell concentration (such thresholding is exactly the operation that we perform during the Bayesian inference to relate tumor concentration with MRI signals, Eq. \ref{eq_mri}). The distributions are centered close to the DICE 1.0. The smaller peak at DICE 0.0 is due to the fact that thresholding of some simulations results in volumes containing a few non-zero voxels, whereas thresholding of the corresponding network predictions outputs volumes of all-zero voxels (or vice versa). In our particular application of the tumor model personalization such volumes are significantly smaller that the binary MRI segmentation volumes to which we calibrate the model and thus do not affect the Bayesian inference outcome. In sum, given that the performance of two different tumor models with notably different diffusion character stays within 2$\%$, we conclude that not only effects imposed by the patient specific CSF anatomy, but also higher order dependency on the WM and GM (through which the diffusion tensor is defined) are captured by the proposed surrogate.

\begin{figure*}[h]
\includegraphics[width=\textwidth]{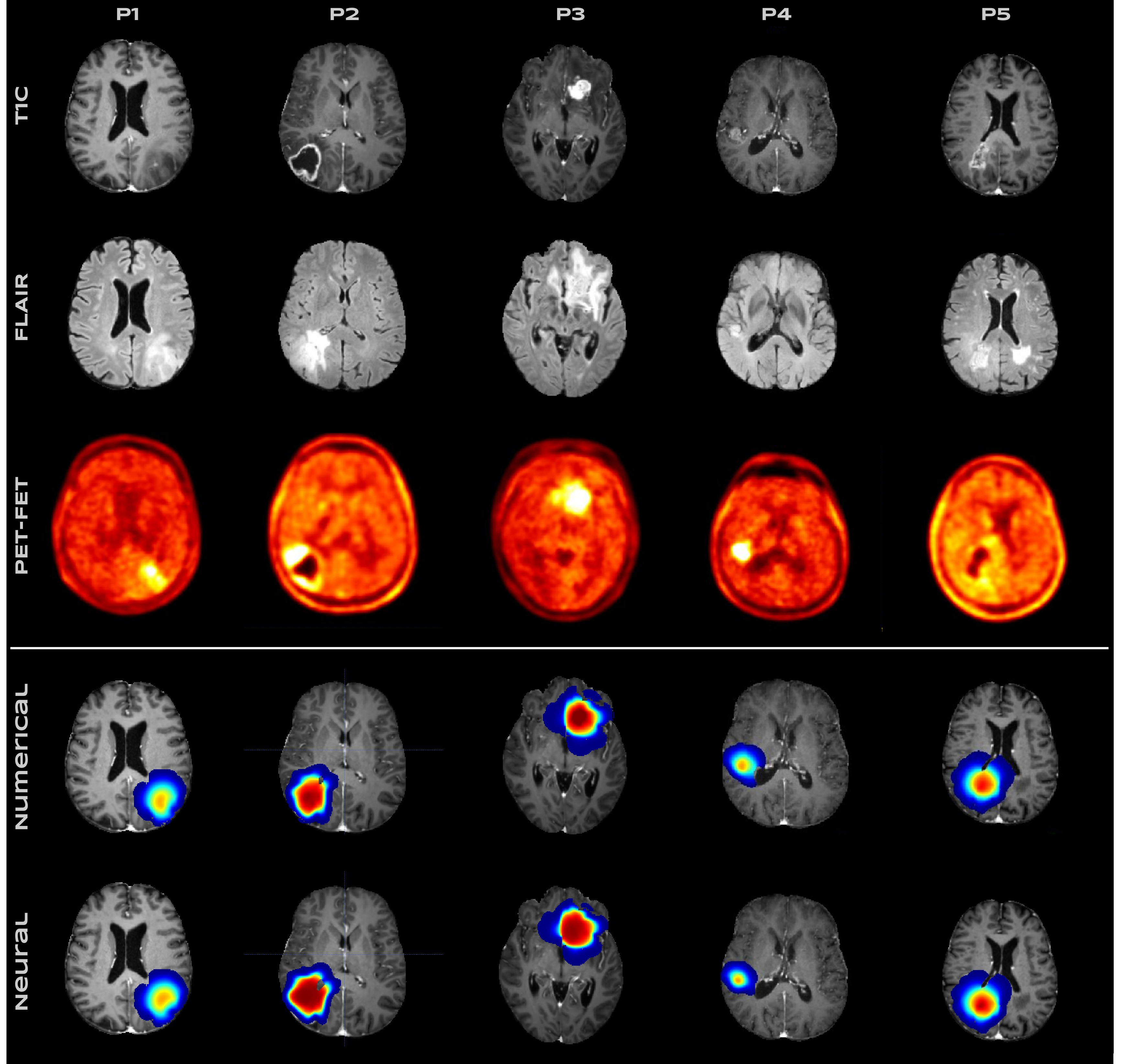}
\caption{Results of the Bayesian inference for patients P1-P5. The three upper rows correspond to the imaging modalities used for tumor model calibration, namely T1c, FLAIR, and FET-PET. The two bottom rows show the simulations of glioma with model parameters inferred via the Bayesian inference using the numerical solver and proposed neural surrogate.} \label{fig7}
\end{figure*}

Fig. \ref{timeerrorfig} depicts the dependency of the mean absolute error versus a network's input - tumor age, showcasing somewhat expected error increase with the increase of the tumor size. 
\begin{table*}
\rowcolors{2}{White}{LightBlue!30}
\centering
\begin{tabular}{*{14}c}\toprule
 & & \textbf{$D$} & \textbf{$\rho$} & \textbf{$T$}  & \textbf{$x$} & \textbf{$y$} & \textbf{$z$}  & \textbf{$\sigma$} & \textbf{$b$} & \textbf{$u^{T1c}_c$}  & \textbf{$u^{FLAIR}_c$} & \textbf{$\sigma_{\alpha}$} & \textbf{MAE} \\\midrule
P1 & \begin{tabular}[c]{@{}c@{}}Numerical\\ solver\end{tabular} & 0.08994 & 0.01550 & 572.582 & 0.3098 & 0.6748 & 0.2819 & 0.2177 & 0.6384 & 0.7976 & 0.3718 & 0.0514  & \cellcolor{White}\multirow{3}{*}{0.09} \\
& \begin{tabular}[c]{@{}c@{}}Neural\\ surrogate\end{tabular} & 0.08976 & 0.01657 & 553.019 & 0.3071  & 0.6657 & 0.2818 & 0.1702 & 0.7198 & 0.7344 & 0.3364 & 0.0500  \\
\hline
P2 & \begin{tabular}[c]{@{}c@{}}Numerical\\ solver\end{tabular} & 0.08920 & 0.01993 & 503.146 & 0.6121 & 0.6706 & 0.3411 & 0.2358 & 0.7309 & 0.6502 & 0.4376 & 0.0688  & \cellcolor{White}\multirow{3}{*}{0.18}  \\
& \begin{tabular}[c]{@{}c@{}}Neural\\ surrogate\end{tabular} & 0.08945 & 0.02882 & 418.631 & 0.6217 & 0.6445 & 0.3316 & 0.2434 & 0.6324 & 0.6316 & 0.4337 & 0.0799 \\
\hline
P3 & \begin{tabular}[c]{@{}c@{}}Numerical\\ solver\end{tabular} & 0.08998 & 0.01555 & 572.128 & 0.4129 & 0.3544 & 0.2900 & 0.2326 & 0.7023 & 0.6088 & 0.3608 & 0.0735  & \cellcolor{White}\multirow{3}{*}{0.15} \\
& \begin{tabular}[c]{@{}c@{}}Neural\\ surrogate\end{tabular} & 0.08990 & 0.02897 & 418.946 & 0.4230 & 0.3678 & 0.2901 & 0.2498 & 0.6561 & 0.6294 & 0.3662 & 0.0772  \\
\hline
P4 & \begin{tabular}[c]{@{}c@{}}Numerical\\ solver\end{tabular} & 0.07967 & 0.00813 & 733.13 & 0.6421 & 0.5502 & 0.3151 & 0.1832 & 1.0146 & 0.6007 & 0.5458 & 0.0757  & \cellcolor{White}\multirow{3}{*}{0.06} \\
& \begin{tabular}[c]{@{}c@{}}Neural\\ surrogate\end{tabular} & 0.04362 & 0.00817 & 776.41 & 0.6707 & 0.5613 & 0.3224 & 0.1966 & 1.0136 & 0.6237 & 0.5192 & 0.0579 \\
\hline
P5 & \begin{tabular}[c]{@{}c@{}}Numerical\\ solver\end{tabular} & 0.08938 & 0.01080 & 686.923 & 0.5530 & 0.6203 & 0.3147 & 0.2484 & 0.6000 & 0.6023  & 0.3775  & 0.0796  & \cellcolor{White}\multirow{3}{*}{0.05} \\
& \begin{tabular}[c]{@{}c@{}}Neural\\ surrogate\end{tabular} & 0.08360 & 0.01066 & 687.841 & 0.5701 & 0.6259  & 0.3276  & 0.2411 & 0.7044 & 0.6098 & 0.3790 & 0.0759  \\
   \bottomrule
 \end{tabular}
 \caption{MAP estimates of the
 marginal distribution from the Bayesian calibration on the patient data using the numerical solver and proposed neural surrogate. The prior ranges are chosen as follows: $D_w \in [0.01, 0.08]$ $mm^2 /day$, $\rho \in [0.0001, 0.03]$ $1/day$, $T \in [30, 1000]$ $days$, $\sigma \in [0.01, 0.25]$, $b \in [0.6, 1.02]$, $u^{T1c}_c \in [0.6, 0.8]$, $u^{FLAIR}_c \in [0.05, 0.6]$, and $\sigma_{\alpha} \in [0.05, 0.08]$. The MAE between the predicted and simulated tumors is computed analogous the 1st term of the network loss - in the area where the tumor from the numerical solver is greater than zero. We observe an agreement of the model parameters obtained by the two methods and the variability of the estimations is within the variability of the Bayesian calibration.}
\end{table*}
Fig. \ref{fig2} qualitatively illustrates the surrogate's predictions over tumor age. 

\subsubsection{Independent testing on the BraTS dataset}
As mentioned above, to provide greater anatomical variability to the training set we performed the training on 64x64x64 crops randomly sampled over the brain volumes of the 10 patients (20k samples in total). That means we heavily randomize over the field of view, inducing variability of the simulation domain that is much larger than what one would see from inter-patient variability (even on the large datasets like BraTS). Thus we argue that training on 20k samples of such data is sufficient for generalisability to an arbitrary test set generated by the same solver within the fixed parametric ranges.
To examine our point, we performed an independent test of the proposed network's generalization performance on a large subset from BraTS composed of all CBICA patients. This test, Fig. \ref{figbrats}, demonstrates generalization capacity that is stable across a high number of tumor and patient anatomies.

\subsubsection{Bayesian model personalization in patient data} As a final test of the surrogate, we performed the Bayesian brain tumor model calibration. We ran the inference on preoperative scans of 5 patients from the in-house validation set using the proposed neural surrogate and numerical solver. The max-a-posterior (MAP) estimates of the tumor density are provided in Tab.2 and tumor concentrations modeled with the MAP estimates are shown in Fig. \ref{fig7}. We observe an agreement of the glioma profiles obtained by the two methods and the variability of the estimations is within the variability of the Bayesian calibration. 

We also observe that the learnable surrogate is trained on a dataset with continuous uniform distribution of the diffusion coefficient $D$ and the proliferation rate $\rho$, whereas the $T$ parameter has a discrete interval. However, during the model calibration, the time parameter is sampled from a continuous interval. This implicitly suggests that the network's interpolation capacity is sufficient for parametric inference. 

\subsubsection{Performance speed-up} 
The computing time for a single simulation using the numerical solver on 64x64x64 grid is on average 15 seconds using an Intel Xeon with 8 CPU cores and 64GB of RAM. The time required by using the neural solver for processing a batch of size 8 during inference is equal to 2.4s, i.e. our surrogate is 50 times faster than the numerical solver per single simulation. Such a speed-up proportionally reduces the computing time for the model personalization.

We want to point out that a theoretical comparison between CPU- and GPU-based computations should be taken with care for a few reasons. First, the neural surrogate does not need to do any sequential computations and can batch process multiple frames, whereas the numerical solver is successively integrating the PDE equation in time and for a single sample. Second, the convolutional layers are very efficiently realized in most GPU-tailored deep learning frameworks. In this regard, a fair comparison could imply to contrast the GPU compute with a CPU cluster containing as many processing units as the GPU. However, the latter is rarely available for practical applications, and as our main goal is clinical translation, we compare CPU and GPU platforms that are widely available. Also, it is worth to mention that ideally the comparison should be with a numerical solver optimized to run on GPUs. 

\section{Discussion}

As depicted in Fig. \ref{fig11}, even though the Dice distributions are peaked close to a perfect score, there is room for improvement. For example, the anatomy encoder well captures the global growth constraints imposed by the CSF anatomy, but tiny CSF components of a few voxels size are often missing in the network predictions, Fig. \ref{fig2}. From our experiments we observe, that the U-Net type model better preserves the CSF, Fig. \ref{unet}. However, in the rest of the volume, the U-Net outputs mispredictions of two types: a) better CSF preservation is traded off for worse capturing of the parametric dependency, b) pronounced false predictions in out-of-tumor area. This leads to a significant performance decrease, Tab. 1. The U-Net is known to work well on datasets which preserve voxel-wise semantics between the network’s input and output \cite{ronneberger2015u,unet1,unet2,unet3}. In our case, though, two out of three input anatomies (WM, GM) do not affect the output (simulated 3D tumor) in a voxel-wise fashion, but rather in a highly non-local way the whole 3D tumor volume. Arguably, the mispredictions (at least type “a”) may be explained by the higher emphasis on the translation of the anatomy input signal  via skip-connections which makes the mapping from the tissues dominate the second input-output mapping - from parameters ($D, \rho, T$) to the tumor volumes. A more intelligent separation of the input-output mappings or hybrid approaches \cite{sht2019implicit} explicitly enforcing boundary conditions might alleviate such issues.  

Another aspect we want to point out is that we do not empirically observe pronounced false predictions in the non-CSF area, even without explicit penalization, as in the proposed loss definition. Moreover, penalization for false predictions in the loss does not improve performance (Tab. 1). First, we attribute the superiority of the loss with only two compartments to the fact that adding extra compartments to the loss introduces a "conflict" of gradients - the optimization focus distorts from the tumor and CSF areas, which are the only areas that affect the numerical simulations. Regarding the false predictions, we can say the following: In tasks like semantic segmentation, a network that is trained without background penalization in the loss would start after some epochs predicting all ones (for a two class problem). This is because the network is incentivized to only push predictions closer to one in the foreground area. If the foreground over the dataset sufficiently covers the whole volume, then the sure-shot solution for the network is to predict ones in the whole volume. In contrast, our tumor simulations dataset contains the foreground that is non-binary (so all ones or all other constant is not an obvious solution), and is in the center of the volume as a single-connected component (so the output volume is not fully covered with foreground). Apparently, a carefully designed network trained on such a dataset is not incentivized to produce anything in the non-penalized area, but low-level noise as a sure-shot, see Fig. \ref{unet}b (arguably, because the further from the center of volume, the lower the range of tumor concentration values the network is exposed to). At the same time, as Fig. \ref{unet}b also shows, using penalization for the tumor complement area does not warrant absence of such noise. Note also, this noise should not affect the Bayesian inference since we threshold our volumes at a higher level when we match the simulated tumors to the real observation according to Eq. 5 and 6.

Lastly, we do not have a definite answer to whether the mentioned limitations and resulting approximation errors are acceptable for clinical translation. Neither do we know whether the Bayesian calibration itself under the simplistic Fisher-Kolmogorov formalism is suitable for the translation. Both require a study on a large cohort of patients, post-surgery analysis, etc. \cite{ref_article3,ref_article7,ref_article11,ender_person_appl}.  However, we find the proposed method to be a solid baseline in the search for optimal tumor model surrogate, which in turn can significantly speed up our search for a biophysical model descriptive enough for clinical trials.


